\documentclass[10pt,letterpaper]{article}
\usepackage[top=0.85in,left=2.75in,footskip=0.75in]{geometry}

\usepackage{amsmath,amssymb}

\usepackage{textcomp,marvosym}

\usepackage{cite}

\usepackage{nameref,hyperref}

\usepackage[right]{lineno}

\usepackage{microtype}

\usepackage[table]{xcolor}

\usepackage{array}

\newcolumntype{+}{!{\vrule width 2pt}}

\newlength\savedwidth

\raggedright
\usepackage[aboveskip=1pt,labelfont=bf,labelsep=period,justification=raggedright,singlelinecheck=off]{caption}

\makeatletter
\renewcommand{\@biblabel}[1]{\quad#1.}
\makeatother

\usepackage{lastpage,fancyhdr,graphicx}
\usepackage{epstopdf}
\fancyheadoffset[L]{2.25in}
\fancyfootoffset[L]{2.25in}
\begin{document}
\vspace*{0.2in}

\begin{flushleft}
{\Large
\textbf\newline{Horizontal transfer between loose compartments stabilizes replication of fragmented ribozymes} 
}
\newline
\\
Atsushi Kamimura\textsuperscript{1*},
Yoshiya J. Matsubara\textsuperscript{1},
Kunihiko Kaneko\textsuperscript{1,2*},
Nobuto Takeuchi\textsuperscript{2,3*}
\\
\bigskip
\textbf{1} Department of Basic Science, Graduate School of Arts and Sciences, The University of Tokyo, 3-8-1, Komaba, Meguro-ku, Tokyo 153-8902 Japan
\\
\textbf{2} Research Center for Complex Systems Biology, Graduate School of Arts and Sciences, The University of Tokyo, 3-8-1, Komaba, Meguro-ku, Tokyo 153-8902 Japan
\\
\textbf{3} School of Biological Sciences, The University of Auckland, Private Bag 92019, Auckland 1142, New Zealand
\\
\bigskip

%
%





* kamimura@complex.c.u-tokyo.ac.jp(AK), kaneko@complex.c.u-tokyo.ac.jp(KK), nobuto.takeuchi@auckland.ac.nz(NT)

\end{flushleft}
\section*{Abstract}
The emergence of replicases that can replicate themselves is a central issue in the origin of life. Recent experiments suggest that such replicases can be realized if an RNA polymerase ribozyme is divided into fragments short enough to be replicable by the ribozyme and if these fragments self-assemble into a functional ribozyme. However, the continued self-replication of such replicases requires that the production of every essential fragment be balanced and sustained. Here, we use mathematical modeling to investigate whether and under what conditions fragmented replicases achieve continued self-replication. We first show that under a simple batch condition, the replicases fail to display continued self-replication owing to positive feedback inherent in these replicases. This positive feedback inevitably biases replication toward a subset of fragments, so that the replicases eventually fail to sustain the production of all essential fragments. We then show that this inherent instability can be resolved by small rates of random content exchange between loose compartments (i.e., horizontal transfer). In this case, the balanced production of all fragments is achieved through negative frequency-dependent selection operating in the population dynamics of compartments. This selection mechanism arises from an interaction mediated by horizontal transfer between intracellular and intercellular symmetry breaking. The horizontal transfer also ensures the presence of all essential fragments in each compartment, sustaining self-replication. Taken together, our results underline compartmentalization and horizontal transfer in the origin of the first self-replicating replicases.

\section*{Author summary}
How evolution got started is a crucial question in the origin of life. One possibility is that RNA molecules gained the ability to catalyze self-replication. Researchers recently proposed how this possibility might have been realized: a long RNA catalyst was divided into short replicable fragments, and these fragments self-assembled into the original long catalyst. Ingenious though it is, we expose a hidden flaw in this proposal.  An auto-catalytic system based on fragmented catalysts involves positive feedback, which necessarily biases replication toward specific fragments and eventually halts the replication of the whole system. However, we also propose an effective remedy to this flaw: compartmentalization and content exchange among compartments generate negative feedback, which tightly coordinates the replication of different fragments.



\section*{Introduction}

One of the crucial questions in the origin of life is how
molecules acquired the capability of undergoing open-ended Darwinian
evolution \cite{joyce2002antiquity, higgs2015rna}. A potential answer is
offered by the template-directed self-replication of a replicase, a
replicase that can replicate itself. To determine whether such
self-replication is demonstrable in RNA, considerable effort has been
devoted to the artificial evolution of RNA polymerase ribozymes
\cite{Johnston2001, Zaher2007, Wochner2011, Attwater2013,
  mutschler2015freeze, Horning2016, 2017OriginsJoyce, Attwater2018}. A
recent milestone in this effort is the demonstration of `riboPCR,' the
exponential amplification of RNA through a PCR-like mechanism catalyzed
entirely by RNA \cite{Horning2016}. The glaring issue, however, has been
that the replicases synthesized so far have limitations in processivity
and fidelity, so that they can replicate only oligomers much shorter
than themselves (or long unstructured cytidine-rich polymers, which
exclude the ribozymes themselves).

As a potential solution to this problem, Mutschler et al.\ and Horning
et al.\ have recently proposed the fragmentation and self-assembly of a
replicase. According to their proposals, a replicase is fragmented into
multiple sequences that are short enough to be replicable by the
replicase and, moreover, capable of self-assembling into a functional
replicase \cite{mutschler2015freeze, 2017OriginsJoyce}. The possibility
of reconstituting a functional ribozyme from its fragments through
self-assembly has been experimentally demonstrated
\cite{mutschler2015freeze, 2017OriginsJoyce, Attwater2018}, attesting
the chemical plausibility of the proposals.

However, the exponential amplification of multiple distinct fragments
raises a question about the dynamical stability of the proposed
autocatalytic system. The continued replication of fragmented replicases
requires the sustained production of all its essential fragments in
yields proportional to the stoichiometric ratio of the fragments in a
replicase \cite{segre2000compositional, furusawa2003zipf,
  Kamimura2018}. However, each fragment is replicated by the replicase
and thus grows exponentially. If some fragments was replicated
persistently faster than the others, the former would out-compete the
latter, causing a loss of some essential fragments and hence the
cessation of self-replication

The above consideration led us to examine whether and under what
conditions fragmented replicases achieve continued
self-replication. Using mathematical modeling, we discovered that the
fragmented replicases fail to display continued self-replication under a
simple batch condition. Replication is inevitably biased toward a subset
of the fragments owing to positive feedback inherent in the replication
cycle of the fragmented replicases, and the loss of fragment diversity
eventually halts self-replication.

To find a way to resolve the above instability, we next examined the
role of compartmentalization. Our model assumes a population of
protocells (primitive cells; hereinafter referred to as ``cells''), each encapsulating a finite number of
fragments and replicases. We found that compartmentalization, in
principle, allows the continued self-replication of the replicases by
the stochastic correction mechanism \cite{szathmary1987group,
  szathmary1995major}. This mechanism is, however, severely limited in
its applicability by its strict requirements on the number of fragments
per cell and the population size of cells. Moreover, this
mechanism is inefficient because it necessitates the continuous
discarding of cells lacking some essential fragments and,
therewith, a large number of fragments that could have produced
functional replicases if combined across discarded cells.

Finally, we show that horizontal transfer between cells provides an
efficient and essential mechanism for the continued replication of the
fragmented ribozymes. The previous studies on imperfect
compartmentalization indicate that such horizontal transfer impedes the
stochastic-correction mechanism \cite{Fontanari2014,
  PhysRevLett.120.158101}. Therefore, horizontal transfer might be
expected to be detrimental to the continued self-replication of the
fragmented replicases. On the contrary, we found that the horizontal
transfer of intermediate frequencies substantially stabilizes the system
to such an extent that the parameter constraints imposed by the
stochastic-correction mechanism are almost completely removed. This
stabilization is caused by negative frequency-dependent selection, which
arises through the interaction between the two distinct types of
symmetry breaking, symmetry breaking between cells and symmetry
breaking within cells, mediated by horizontal transfer.

\section*{Model}

We consider the simplest model of fragmented replicases, in which a
catalyst consists of two fragments.  The fragments (denoted by $X$ and
$Y$) self-assemble into the catalyst (denoted by $C$), and the catalyst
disassembles into the fragments as follows:
\begin{align}
X + Y \xrightarrow{k^f} C, \hspace{5mm} C \xrightarrow{k^b} X + Y.
\label{Reaction1}
\end{align}
We assume that the catalytst cannot replicate its own copies, but can
replicate its fragments because shorter templates are more amenable to
ribozyme-catalyzed replication as mentioned above. Therefore,
\begin{align}
X + C \xrightarrow{k_x} 2X + C,\hspace{2mm} Y + C \xrightarrow{k_y} 2Y + C,
\label{Reaction2}
\end{align}
where the monomers are ignored under the assumption that their
concentrations are buffered at constant levels, and complementary
replication is ignored for simplicity. In the presence of the catalyst,
each fragment replicates at a rate proportional to its copy
number. Hence, the fragments undergo exponential amplification.

\section*{Results}

\subsection*{Failure of balanced replication of fragments under a batch
  condition}

First, we show that the replication of the fragments $X$ and $Y$ are
unstable in a batch condition: replication is biased toward either of
the fragments even if the rate constants for $X$ and $Y$ are identical,
and the minor fragment is gradually diluted out from the system, so that
the replication of the catalysts eventually stops. In this paper, we
mainly focus on the situation where the rate constants are equal
($k_x = k_y = k$) because our results remain
  qualitatively the same as long as the difference between $k_x$ and
  $k_y$ is sufficiently small.

We assume that the reactions undergo in a well-mixed batch condition so that 
the dynamics of the concentrations of $X$, $Y$, and $C$ (denoted by $x$,
$y$, and $c$, respectively) are written as follows:
\begin{align}
\frac{dx}{dt} &= \left( - k^f xy + k^b c + k xc \right) - x \phi 
\label{xdynamics}
\\
\frac{dy}{dt} &= \left( - k^f yx + k^b c + k yc \right) - y \phi 
\label{ydynamics}
\\
\frac{dc}{dt} &= \left( k^f xy - k^b c \right) - c \phi, 
\label{cdynamics}
\end{align}
where $\phi = k(x + y) c$.  In the right-hand side of these equations,
the first terms in the brackets represent chemical reactions, and the
second terms multiplied by $\phi$ represent dilution. This dilution terms
are introduced to fix the total concentration of fragments, $x+y+2c$,
since $x$ and $y$ increase through replication. Within the brackets
enclosing the reaction terms, the first and second terms represent
forward and backward reactions of \ref{Reaction1}, respectively.  The
third terms, which are present only in Equations \ref{xdynamics} and \ref{ydynamics}, denote the replication of $X$ and $Y$ through
reactions \ref{Reaction2}, respectively.

By introducing variables $x_{tot} = x + c$ and $y_{tot} = y + c$, one
can write
\begin{align}
\frac{d}{dt} \left( \frac{x_{tot}}{y_{tot}} \right) = \frac{kc^2}{y^2_{tot}} \left( x_{tot} - y_{tot} \right).
\end{align}
This equation indicates that a steady-state solution satisfies
$x_{tot} = y_{tot}$.  This solution is, however, unstable: a small
increase in, say, $x_{tot}$ over $y_{tot}$ gets amplified because
$kc^2/y^2_{tot}$ is always positive, and, as a consequence, replication
is biased to $X$.  Intuitively, when $x_{tot}$ is slightly greater than
$y_{tot}$, the amount of free fragments $x$ must also be greater than
$y$ because the same amount of $X$ and $Y$ are incorporated into
catalysts.  Therefore, the replication of
$X$ occurs more frequently than that of $Y$ because more templates of
$X$ are available.  As a result, the increase of $x_{tot}$ is greater
than that of $y_{tot}$. Because of this positive feedback, the
concentration of the minor fragment $Y$ gradually decreases as it is
diluted out from the system, and, as a consequence, that of the
catalysts $C$ also decreases.  Finally, the replication reaction stops
once the catalysts are lost from the system.

The instability of replication under a batch condition can be generally
demonstrated for catalysts composed of an arbitrary number of fragments
by straightforwardly extending the above model [see Supporting text section 1].

\subsection*{Compartmentalization can overcome the unstable replication
  by selecting out non-growing cells but only under strong constraints on
  the sizes of cell volume and population}

The introduction of compartments and their competitions can overcome the
unstable replication.  When the system is compartmentalized into a
number of cells, stochasticity in cell compositions, competition for
growth and division of cells provide a possible solution to avoid the
loss of fragments: As the cells grow and eventually split into two with
fragments distributed randomly between the daughter cells, cells with
both $X$ and $Y$ fragments continue growth, while cells without either
of them cannot grow.  By introducing such a stochastic-correction
mechanism at the cell level \cite{szathmary1987group}, one expects that
the instability by the positive feedback at the molecule level can be
resolved.  To investigate this, we assume that the fragments and their
assembly to function as a catalyst are partitioned into $N_{cell}$
cells: the reactions occur in each cell.
We adopted stochastic simulation using Gillespie
algorithm \cite{Gillespie} for reactions
\ref{Reaction1} and \ref{Reaction2}.  We assume that the volume of each
cell is proportional to the number of fragments inside, and as the
number of fragments increases in a cell, the cell grows.  When the total
number of fragments reaches a threshold $V_{Div}$, the cell divides with
the components randomly partitioned into two daughter cells.  Here, at
the division event, one randomly-chosen cell is removed to fix the total
number of cells $N_{cell}$.  By this cell-cell competition, cells with
biased composition of $X$ and $Y$ are selected out because their growth
is slow.

\begin{figure}
\begin{center}
\includegraphics[width=\linewidth]{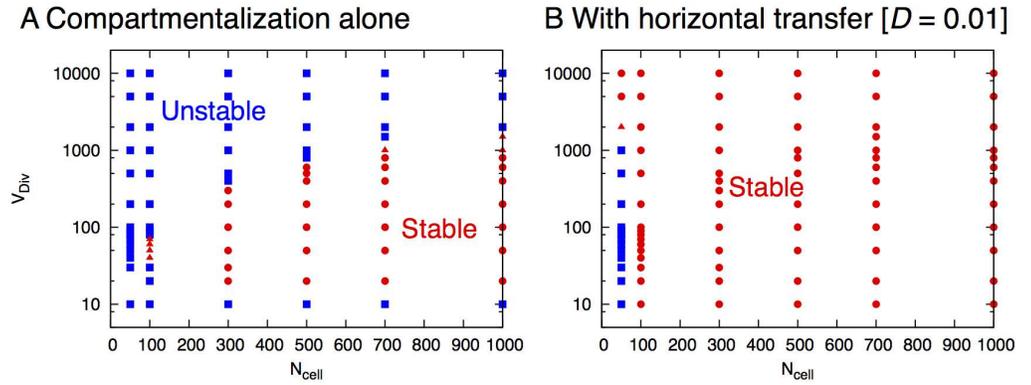}
\caption{{\bf Sets of division threshold $V_{Div}$ and the number of cells $N_{cell}$ with which the unstable replication of reactions \ref{Reaction1} and \ref{Reaction2} are avoided by (A) compartmentalization alone and (B) that with horizontal transfer of the transfer constant $D = 0.01$.}
For the sets shown as stable [red circles], the system can continuously have cells with both fragments in the simulations up to $4\times 10^5$ division events from an initial condition where $V_{Div}/4$ copies of each $X$ and $Y$ are in each cell.
For the sets shown as unstable [blue squares], all cells with both fragments are lost from the system and it cannot continue growth. 
For the sets located at the boundary of stable and unstable area [shown in red triangles], the outcome depends on simulation runs. }
\label{fig:1}
\end{center}
\end{figure}

The relevant parameters for controlling the effect of compartmentalization are the division threshold $V_{Div}$ and the number of cells $N_{cell}$.
Figure \ref{fig:1}A shows sets of the parameters with which the stochastic-correction mechanism can avoid the unstable replication, by suppressing the positive feedback and 
selecting cells keeping both fragments.  
If $V_{Div}$ is very small (of the order of $10$), the stochasticity of cell components is too strong to maintain both fragments continuously and either of them is lost for all cells. 
Hence, the system cannot continue growth. 
On the other hand, if $V_{Div}$ is too large, stochasticity in components decreases. 
In each cell, the balance of fragments is broken, and the replication is biased to either of $X$ or $Y$.
Then, components of each cell are dominated by either of free $X$ or $Y$, and 
the number of catalysts in dividing cells gradually decreases to one because at least one catalyst is necessary to replicate fragments.
Even when the $N_{cell}$ cells are separated into the equal number of $X$-dominant and $Y$-dominant cells,  
there is no frequency-dependent selection between the cells. 
Thus, the random drift will finally result in bias to either of $X$-dominant or $Y$-dominant cells.
By division events, daughter cells without catalysts randomly replace remaining cells, therefore, the cells with catalysts are finally removed from the system.  

For values of $V_{Div}$ in-between, some of $N_{cell}$ cells keep both $X$ and $Y$, and can continue the replication. 
Besides $V_{Div}$, the number of cells $N_{cell}$ is also restricted, to maintain such cells keeping both fragments $X$ and $Y$.
At division events, dividing cells without both fragments randomly replace remaining cells. Hence, when the number of cells $N_{cell}$ is small, all the cells with both fragments will be finally removed.  
As the number of cells $N_{cell}$ increases, the probability decreases that all the cells with both fragments are 
removed. As a result, the range of $V_{Div}$ with stable replication increases.
Note that the above cell-level selection mechanism is based on the removal of non-growing cells, and is inefficient because a large number of fragments in cells lacking some fragments must be continuously removed from the system although they are still functional if combined across the non-growing cells. 

\subsection*{Horizontal transfer of fragments with small rates removes the constraints of compartments for stable replication}

Without the selection in cell population nor the restriction to $V_{Div}$, 
horizontal transfer of fragments between cells rescues the loss of fragments and enables continuous replication by maintaining the balance between $X$ and $Y$.  
If the $X$-dominant and $Y$-dominant cells coexist in the cell population, 
the transfer between cells avoids loss of fragments for both cells by supplying fragments to each other because each fragment is in excess for cells on one side 
but lacking for cells on the other side.

For the purpose, we consider random mutual transfers of molecules among the $N_{cell}$ cells. 
To implement the transfer, we consider reactions, $X \xrightarrow{D} 0$, $Y \xrightarrow{D} 0$, $C \xrightarrow{D} 0$
so that the $X$, $Y$ and $C$ are removed from a cell, respectively, with rate in proportional to each concentration, i.e., $Dx$, $Dy$, and $Dc$. 
This gives diffusion out of the cell. 
At the same time, the component is added to another randomly-chosen cell. 

\begin{figure}
\begin{center}
\includegraphics[width=\textwidth]{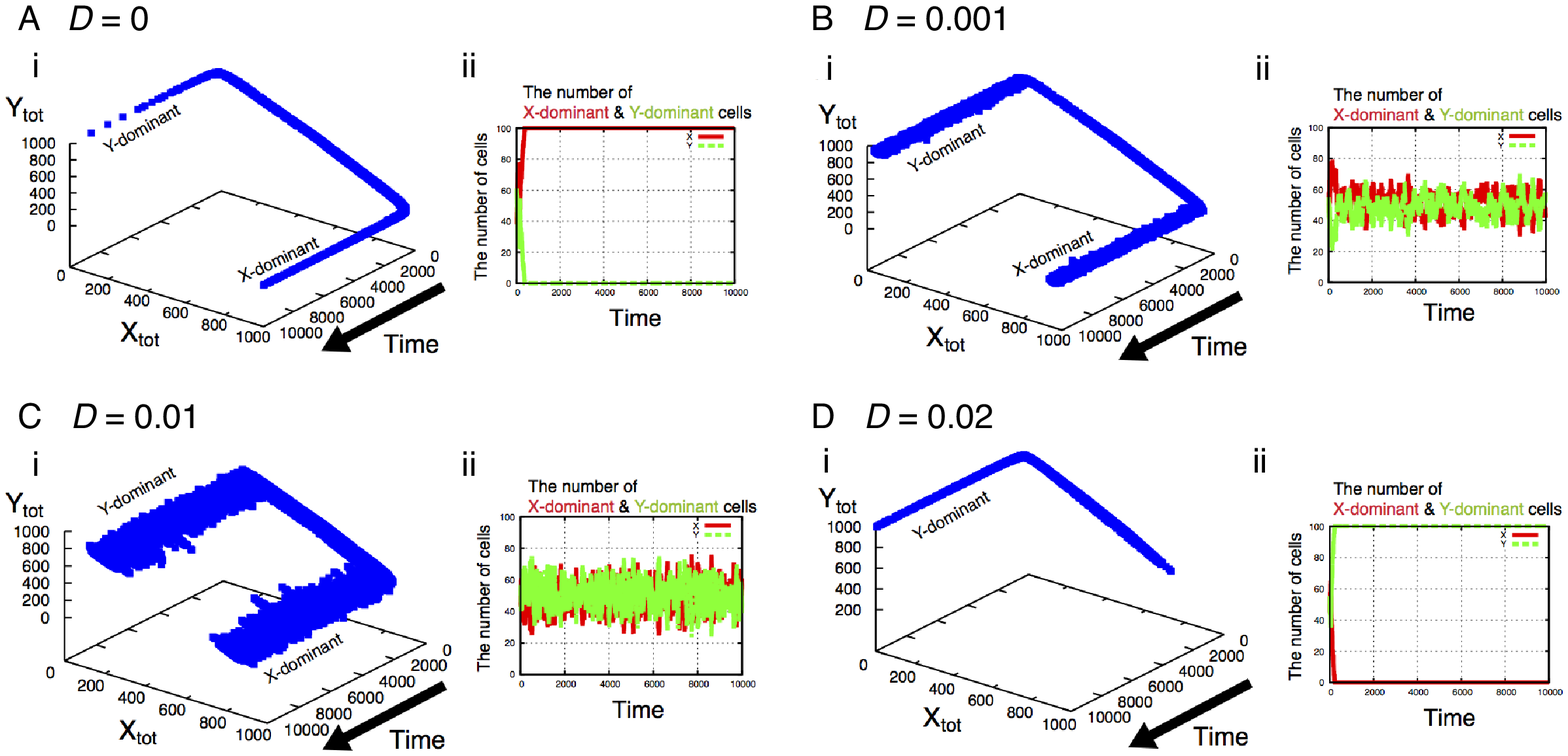}
\caption{
{\bf (i) The number of fragments $X_{tot}$ and $Y_{tot}$ of dividing cells and (ii) the number of $X$-dominant and $Y$-dominant cells at corresponding time for the transfer rates (A) $D = 0$ (B) $D = 0.001$ (C) $D = 0.01$ and (D) $D = 0.02$.}
Initially, the numbers of $X_{tot}$ and $Y_{tot}$ are approximately equal and, as time goes on, 
cells are differentiated into either of $X$-dominant or $Y$-dominant compositions. 
For $D=0$ (A), the system is unstable: only $X$-dominant cells (for this run) dominate (ii) and finally, cells cannot continue growth.
For $D = 0.001$ (B) and $0.01$ (C), the system is stable; $X$ and $Y$ fragments coexist in each cell with unequal population (i).
Here, the asymmetry between the major and minor fragments gets smaller as $D$ increases. In addition, the two types of cells, $X$-dominant and $Y$-dominant 
cells coexist with the equal population (ii). As $D$ increases further [$D = 0.02$ (D)], the system gets unstable and only either of $X$ or $Y$ remains (ii).
The parameters are $V_{Div} = 1000$, $N_{cell} = 100$, $k^f = f^b = 1$, $k_x = k_y = 1$.
}
\label{fig:2}
\end{center}
\end{figure}

With the transfer among cells, the replication of the fragments is stabilized when the transfer constant $D$ is small.
In fact, the constraints of $V_{Div}$ and $N_{cell}$ are drastically eliminated [Figure \ref{fig:1}B]. As long as the parameters are not extremely small, the stable replication continues. 
For small positive values of $D$ [Figures \ref{fig:2}B and C], the cell keeps on growing with the coexistence of $X$ and $Y$ molecules in each cell, even for large $V_{Div}$ where only $X$-dominant or $Y$-dominant cells remain for $D = 0$ [Figure \ref{fig:2}A].
Here, the asymmetry between the fractions of the major and minor fragments gets smaller as $D$ increases.
In addition, two types of cells, $X$-dominant and $Y$-dominant cells coexist roughly with equal population [Figures \ref{fig:2}B(ii) and C(ii)].
As $D$ is increased further [Figure \ref{fig:2}D], the system gets unstable and only either of $X$ or $Y$ remains.
This is natural, because for a large $D$ limit, the system is well mixed, and the system is reduced back to the case without compartmentalization.

\subsection*{Bifurcation explains the stable replication with small rates of horizontal transfer in two subsystems 
as an approximation of cell population}

To answer why the small rates of transfer stabilizes the system, we approximate the dynamics of the population of cells by considering the dynamics of two subsystems between which the fragments are transferred. 
We assume the equal population of $X$-dominant and $Y$-dominant cells as two subsystems of equal volumes, namely, $1$ and $2$, respectively.
We write the total concentration of $X$ (the total of free $X$s and $C$s) of each subsystem as $x^1_{tot}$ and 
$x^2_{tot}$, and the total of $Y$ as $y^1_{tot}$ and $y^2_{tot}$. 
The dynamics of $x^i_{tot}$ in each subsystem($i=1,2$) is written as 
\begin{align}
\dot{x}^1_{tot} = \frac{dx^1_{tot}}{dt} &= F^1 - \frac{D}{2} x^1_{tot} + \frac{D}{2}x^2_{tot}, 
\label{meanx1}
\end{align}
\begin{align}
\dot{x}^2_{tot} = \frac{dx^2_{tot}}{dt} &= F^2 - \frac{D}{2} x^2_{tot} + \frac{D}{2}x^1_{tot}.
\label{meanx2}
\end{align}
where $F^i = k (x^i_{tot} - c^i)c^i - x^i_{tot} \mu_i$.
The first term in $F^i$ represents the replication of the component $X$ by the first reaction of 
\ref{Reaction2}, where $k$, $(x^i_{tot} - c^i)$, and $c^i$ denote the
rate constant, the concentrations of free $X$ and $C$ in
subsystem $i$, respectively.  
The second term multiplied by $\mu_i$ in $F^i$ represents the dilution effect of the component due to the volume growth of the subsystem. 
The volume growth is assumed to keep the total concentration at unity (i.e., $x^i_{tot} + y^i_{tot} = 1$). 
Accordingly, $\mu_i$ is defined as $\mu_i = k (x^i_{tot} - c^i) c^i + k (y^i_{tot} - c^i) c^i = k ( 1 - 2c^i ) c^i$.
Thus, in each subsystem, the components are diluted by the rate $\mu_i$ in total. Then, the dilution rate of each component is proportional to the amount of the component, therefore, the component $X^i_{tot}$ is diluted by the factor $x^i_{tot} \mu_i$. Along with the volume growth of each subsystem, we also assume that the total volume of both subsystems is fixed identical by removing components of each subsystem in proportion to its volume. 
This process corresponds to the random removal of cells to fix $N_{cell}$ cells in our simulations, and the volume ratio of the subsystem 1 to 2 dynamically changes in general. In this section, we fix the two subsystems with an equal volume, whereas the dynamics of the volumes is investigated in the next section.  
Next, the second and third terms in Equations \ref{meanx1} and \ref{meanx2} denote average out- and in-flow
of the components $X$ by the transfer, respectively.  
These average flows are estimated as follows: 
the amount of the fragment $X$ diffused out from the subsystem $1$ is $Dx^1_{tot}$, 
but half of them is returned to the subsystem itself because, in our simulation, the population of cells is divided into $X$-dominant and $Y$-dominant cells with the equal population of $N_{cell}/2$ and each fragment diffused out from a cell is  randomly re-distributed into one of the cells, i.e., half of the fragments are distributed into $X$-dominant cells.
Thus, the effective amount of fragments diffused out of subsystem $1$ to $2$ is $Dx^1_{tot}/2$.  
In the same manner, the effective amount of the fragment $X$ for subsystem $1$ received from subsystem $2$ is approximated as halves of $Dx^2_{tot}$. 
The dynamics for $Y$ are estimated in the same manner.

\begin{figure}
\begin{center}
\includegraphics[width=\textwidth]{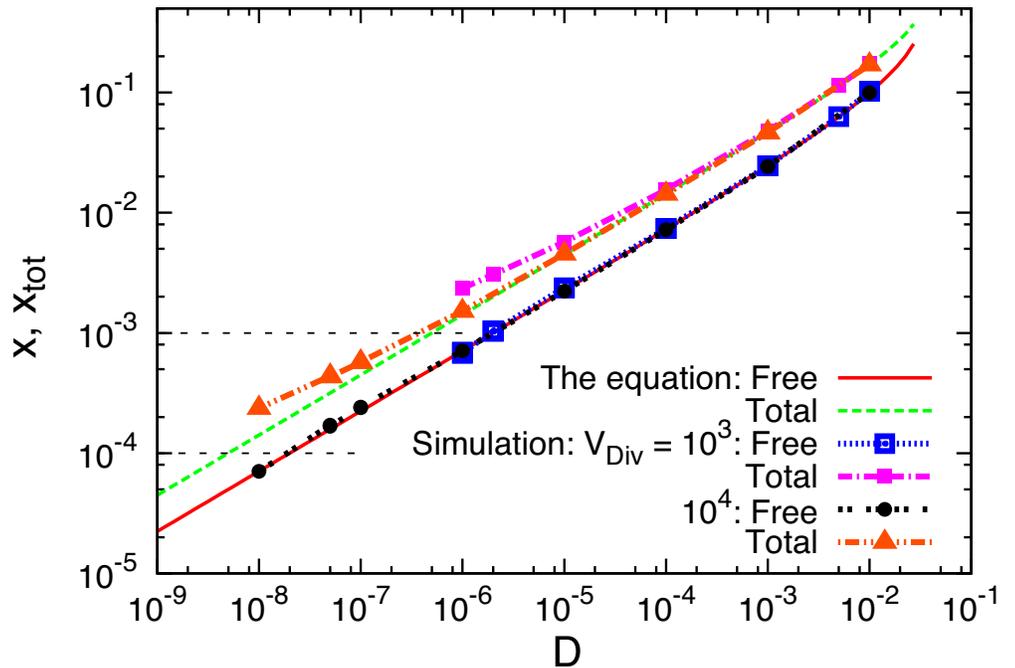}
\caption{{\bf The concentration $x$ and $x_{tot}$ at division events for $Y$-dominant cells as a function of $D$.} Free and Total indicate $x$ and $x_{tot}$, respectively.
For the free fragments [$x$], the results of simulations [Blue and black curves for $V_{Div} = 10^3$ and $10^4$] agree well with the solution of Equation \ref{fixpoint2} [Red curve].
For the total fragments [$x_{tot}$], the simulations [Magenta and orange curves for $V_{Div} = 10^3$ and $10^4$] agree with the solution of Equation \ref{fixpoint2} [Green curve] 
for larger $D$, but shift to larger values for smaller $D$. This is because cells must possess at least one catalyst to divide, therefore, the total fragments including $c$ shift to larger values as it approaches the minimum requirement. 
For reference, the values of $x_{tot} = 1/V_{Div}$ at which the number of $c$ is equal to one for $V_{Div} = 10^3$ and $10^4$ are shown by horizontal dotted lines, respectively. 
}
\label{fig:3}
\end{center}
\end{figure}

The fixed-point solutions of Equation \ref{meanx1} are analytically obtained by approximating  $c^i \approx 1 - \sqrt{1 - x^i_{tot} y^i_{tot}} \approx x^i_{tot} y^i_{tot}/2$. 
The first approximation assumes that the dynamics of reaction \ref{Reaction1} is much faster than
those of reactions \ref{Reaction2} and transfers. 
The concentration of $c^i$ is, then, approximated from the condition $dc^i/dt = 0$ as
$c^i = 1 - \sqrt{1 - x^i_{tot} y^i_{tot}}$ for $k^f = k^b$.
In addition, as we are interested in the stability of the system against the biased replication, we consider the case of highly-asymmetric composition, i.e., $x^i_{tot} y^i_{tot} \ll 1$.  
Then, the second approximation $c^i = x^i_{tot} y^i_{tot} /2$ is obtained by using $(1-\epsilon)^{1/2} = 1-\epsilon/2$ for $|\epsilon| \ll 1$.
Using the approximations, the stable fixed point is obtained as 
\begin{align}
x^1_{tot} = \frac{1}{2} \left( 1 + \sqrt{1 - 4\sqrt{2D/k}} \right),
\label{fixpoint1}
\end{align}
and
\begin{align}
x^2_{tot} = \frac{1}{2} \left( 1 - \sqrt{1 - 4\sqrt{2D/k}} \right).
\label{fixpoint2}
\end{align}
Here, we assume the dominant fragment of subsystem 1 is $X$ and that of subsystem 2 is $Y$ because $x^1_{tot} >1/2$ and $x^2_{tot} <1/2$.
Further, the relative value of the transfer rate $D$ to the replication rate $k$ is essential so that we fix $k=1$ below. 
The solution of the minor fragment $x^2_{tot}$ is plotted as a function of the transfer rate $D$ in Figure \ref{fig:3}.
As for the free fragments $x^2 = x^2_{tot} - c^2$, it agrees well with the results of our simulation.
For the total fragment $x^1_{tot}$, it agrees with the simulations for greater $D$ although the results of simulation shift to larger values for smaller $D$, where the number of fragments decreases, but cells must possess at least one catalyst to divide, so that the total fragments including $c$ shift to greater values as it approaches the minimum requirement, i.e., 
the total concentration of minor fragments be $\geq 1/V_{Div}$.

\begin{figure}
\begin{center}
\includegraphics[width=\textwidth]{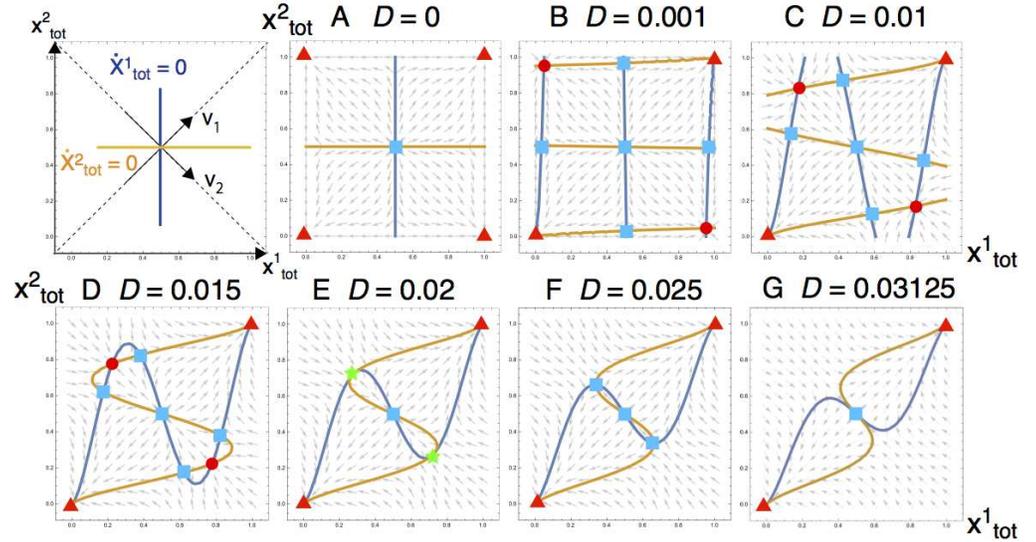}
\caption{{\bf Flow diagram of Equation \ref{meanx1}.}  As schematically indicated in the left-top panel, the nullclines are shown for $\dot{x}^1_{tot} = 0$ and $\dot{x}^2_{tot}= 0$ in blue and orange, respectively, and the crossing points of them are solutions. The directions of $v_1 = (1,1)$ and $v_2 = (1, -1)$ are also indicated. For the solutions, stable fixed points are shown in red: those with stable growth [i.e., both fragments are in each subsystem] are in red circles, and those without growth [either of fragments is lost from subsystems or whole systems] are in red triangles. Unstable solutions are in light-blue squares, and neutral solutions in the $v_1$-direction are in green stars at $D = 0.02$ (E). For $D = 0$ (A), the solution exists at $(x^1_{tot}, x^2_{tot}) = (1/2, 1/2)$ but it is unstable.  For small values of $D$ (B to D), the stable fixed points with growth [red circles] appear in addition to fixed points without growth.  At $D = 0.02$ (E), the fixed points with growth get unstable (shown in green stars) in $v_1$-directions. As $D$ increases further (F), the two fixed points are still stable in $v_2$-directions, while the solution at $(x^1_{tot}, x^2_{tot}) = (1/2, 1/2)$ is unstable in the direction. At $D = 0.03125$ (G), the system transits from the three fixed points to one fixed point. }
\label{fig:4}
\end{center}
\end{figure}

To study further the stability of the solution, we plot the flow [a direction of the vector ($\dot{x}^1_{tot}, \dot{x}^2_{tot}$)] 
of Equation \ref{meanx1} in Figure \ref{fig:4}. 
The steady-state solutions satisfy both $\dot{x}^1_{tot} = 0$ and $\dot{x}^2_{tot} = 0$, therefore, they 
are represented as the crossing points of two nullclines [set of $(x^1_{tot}, x^2_{tot})$ satisfying $\dot{x}^1_{tot} = 0$ or $\dot{x}^2_{tot} = 0$, indicated by blue and orange curves (see left-top panel)].
For $D = 0$ [Figure \ref{fig:4}A], a solution exists at $(x^1_{tot}, x^2_{tot}) = (1/2, 1/2)$ (indicated by the light-blue square).
However, it is unstable because the flows (arrows) point outward from the solution. 
Then, the system moves away from the solution by any tiny perturbation. 
The flows point toward each corner of the plane (indicated by the red triangles), where either of $X$ or $Y$ is lost and cells cannot grow. 
For small positive values of $D$ [Figures \ref{fig:4}B to D], 
stable fixed points (shown in red circles) appear to which the flows are directed from all directions, 
in addition to unstable fixed points (shown in blue squares) and the trivial solutions $(x^1_{tot}, x^2_{tot})=(0,0),(1,1)$ (shown in red triangles).
Note that there exist two stable fixed points [red circles] for each $D$ [Figures \ref{fig:4}B to D], and the solution in Equation \ref{fixpoint1} corresponds to the right-bottom one.

As $D$ increases, a bifurcation occurs at $D = 0.02$ [Figure \ref{fig:4}E] so that the stable fixed points for $D \leq 0.02$ turn to be unstable (green stars). 
To understand this bifurcation, we consider eigenvectors $v_1$, $v_2$ of Jacobian matrix of Equation  \ref{meanx1} for the eigenvalues $\lambda_1$ and $\lambda_2$. 
At the stable fixed points, they are obtained as $v_1 = (1,1)$ and $v_2 = (1,-1)$ [see left-top panel in Figure \ref{fig:4}].
The direction of $v_1$ determines the asymmetry between $X$ and $Y$ in both subsystems.
By moving along the $v_1$-direction of the plane, the amount of $x^1_{tot} + x^2_{tot}$ either increases or decreases 
while $y^1_{tot} + y^2_{tot} = 2 - (x^1_{tot} + x^2_{tot})$ decreases or increases, respectively.
On the other hand, the direction of $v_2$ corresponds to the asymmetry between subsystems $1$ and $2$ for the fragments of $X$.
By moving along the $v_2$-direction of the plane, the amount of $x^1_{tot}$ increases or decreases while $x^2_{tot}$ decreases or increases, respectively. 
The corresponding eigenvalues for $v_1$ and $v_2$ are calculated as 
$\lambda_1 = 5D - \frac{\sqrt{2D}}{2}$ and $\lambda_2 = 4D - \frac{\sqrt{2D}}{2}$, respectively. 
As $D$ increases, a bifurcation occurs first in $v_1$-direction at $D^* = 0.02$ which is obtained from $5D^* - \frac{\sqrt{2D^*}}{2} = 0$.
In fact, the flows (arrows) at the fixed point (green stars) are in the parallel direction of $v_2$, 
and point outward in the $v_1$-directions as $D$ is increased further.
This corresponds to the case in which the symmetry between $X$ and $Y$ breaks and only either of $X$ and $Y$ remains in both systems.
The estimated value of $D^*$ agrees with the results of our simulation [Figure \ref{fig:2}].
In the two subsystems, the bifurcation also occurs in $v_2$-direction at $D^+ = 0.03125$, as obtained from $4D^+ - \frac{\sqrt{2D^+}}{2} = 0$, 
corresponding to the symmetry between subsystems $1$ and $2$.
At the bifurcation point, the three fixed points (one unstable and two stable points in $v_2$-directions; shown all in light-blue squares) 
merge to one fixed point [Figure \ref{fig:4}G].

The behavior of the bifurcations can be understood as follows. 
The system has two kinds of symmetry, one between fragments $X$ and $Y$, and one between subsystems $1$ and $2$. 
For the stable replication, the symmetry between $X$ and $Y$ should be maintained because both fragments are essential. 
On the other hand, the symmetry between subsystems $1$ and $2$ should be broken because each fragment should be in excess for one subsystem, 
but lacking for the other subsystem. The two subsystems `help' each other by the transfer of molecules.
The former symmetry is maintained for $0 \leq D < D^*$ and breaks for $D > D^*$. 
On the other hand, the latter symmetry is broken in the range $0 \leq D < D^+$.
To meet the two conditions for the stable replication, the values of $D$ are restricted as $0 < D < D^* = 0.02$ because $D^* < D^+$
($D = 0$ is eliminated by the condition each subsystem should contain both fragments).

\subsection*{Frequency-dependent selection: why the balance of fragments is achieved at the cell population}

In the previous section, we confirmed the stable replication by small rates of horizontal transfer, by assuming that the populations of two cell types are equal. 
Here, we verify that the state of the equal volume i.e. the equal population of $X-$ and $Y-$dominant cells, 
is stable and selected as a result of a frequency-dependent selection. 
To analytically investigate the stability of the state, we consider the volume fractions of the subsystems $1$ and $2$ are slightly different from 1/2, to be replaced by 
$1/2 + \epsilon$ and $1/2 - \epsilon$, respectively, with $\epsilon$ as a small parameter.
Then the rate equations \ref{meanx1} and \ref{meanx2} are 
\begin{align}
\frac{dx^1_{tot}}{dt} = F^1 - D \left( \frac{1}{2} + \epsilon \right) x^1_{tot} + D \left( \frac{1}{2} - \epsilon \right) x^2_{tot}, 
\label{stabilityFDS1}
\\
\frac{dx^2_{tot}}{dt} = F^2 - D \left( \frac{1}{2} - \epsilon \right) x^2_{tot} + D \left( \frac{1}{2} + \epsilon \right) x^1_{tot},
\label{stabilityFDS2}
\end{align}
where the replication and the dilution terms due to the volume growth are written as $F^i = - x^{i2}_{tot} (1-x^i_{tot})^2 (1 - 2x^i_{tot})/4$ by substituting the approximation $c^i = x^i_{tot} y^i_{tot}/2$.

Below, we show that the growth rate of the minor subsystem 2 with the fraction $1/2 - \epsilon$ (for $\epsilon > 0$) increases and the major subsystem 1 decreases, leading the fraction of the two subsystems to go back to equal.
First, we write the concentrations of $X$ at the steady state as $x^1_{tot} = x^* + \delta_1$ and $x^2_{tot} = 1 - x^* + \delta_2$ where 
$x^* = \frac{1}{2} \left( 1 + \sqrt{1 - 4\sqrt{2D}} \right)$ is the solution for $\epsilon = 0$ (Equation  \ref{fixpoint1}),  
and $\delta_1$ and $\delta_2$ are deviations caused by 
the introduction of $\epsilon$, respectively, for $x^1_{tot}$ and $x^2_{tot}$.
Then, from the steady state condition of Equations \ref{stabilityFDS1} and \ref{stabilityFDS2}, one gets 
$\delta_1 = \delta_2 =  \frac{\sqrt{D} \epsilon}{\sqrt{2}/2 - 4\sqrt{D}}$.
The growth rates $\mu_i$ $(i = 1,2)$ are given by $(x^i_{tot} - c^i) c^i + (y^i_{tot} - c^i) c^i$ so that 
\begin{align}
\mu_1 = \mu^* - \gamma(D) \epsilon, 
\label{growth1}
\\
\mu_2 = \mu^* + \gamma(D) \epsilon,
\label{growth2}
\end{align}
where $\mu^* = \frac{\left\{ 1 - x^* (1-x^*) \right\} x^* (1-x^*) }{2}$ is the growth rate at $\epsilon = 0$,  
and $\gamma(D) =  \frac{1-2\sqrt{2}D}{\sqrt{2}\sqrt{1-4\sqrt{2D}}} \sqrt{D} > 0$, showing the decrease and the increase of the subsystem 1 and 2, respectively by $\epsilon$.

When $\epsilon > 0$, i.e., the volume of subsystem $1$ exceeds that of $2$, 
the concentrations of $X$ in both subsystems $1$ and $2$ increase ($\delta_1 = \delta_2 > 0$).
For the subsystem $1$, the fragment $X$ is majority ($x^1_{tot} = x^* > 1/2$), therefore, the asymmetry between $X$ and $Y$ is enhanced by the increase of $X$.
On the other hand, the fragment $X$ is the minority in subsystem $2$, and the composition of $X$ and $Y$ gets close to be symmetric by $\delta_2$.
Because the growth rate is maximized when the concentrations of $X$ and $Y$ are equal, the rate $\mu_1$ of subsystem $1$ decreases, while that of $2$, $\mu_2$, increases by the factor $\gamma(D) > 0$ (see Equations  \ref{growth1} and \ref{growth2}). 
Consequently, the volume ratio of the two subsystems eventually goes back to equal. 
(Note that the frequency-dependent selection remains for any non-zero $D$, however, the replication is unstable in our simulation for small values of $D$ when discreteness in molecule number is taken into account [see Supporting text section 2]).

\section*{Discussions}

In summary, we have shown that the self-replication of fragmented
replicases is unstable under a simple batch condition. Replication is
biased towards a subset of the fragments and eventually stops due to the
lack of an essential fragment. Although the stochastic-correction
mechanism induced by compartmentalization helps, it imposes severe
restrictions on the number of molecules per cell and the population
size of cells. In addition, the mechanism is inefficient
because a large number of fragments in non-growing cells must be
discarded.  Finally, we have shown that the horizontal transfer of
intermediate frequencies provides an efficient and favorable solution to
the instability of the fragmented replicases.

Recent experimental studies have been challenged to use self-assembling
fragmented ribozymes to synthesize each of the component fragments to
achieve the RNA-catalyzed exponential amplification of the ribozyme
itself \cite{2017OriginsJoyce}. The self-assembly of functional RNA
polymerase ribozymes from short RNA oligomers has been demonstrated
by driving entropically disfavored reactions under iterated freeze-thaw
cycles \cite{mutschler2015freeze}.  Our theoretical results predict that
these approaches for (re-)constructing RNA-based evolving systems have
the serious issue: the replication of fragments is inevitably biased, so
that it eventually fails to produce the copies of the ribozymes.
Simultaneously, our study proposes a solution for this issue:
the random exchange of fragments between loose compartments at intermediate frequencies.

Recent experiments suggest that the random exchange of contents between
compartments is plausible. The freeze-thaw cycles, which enhance the
assembly of fragments \cite{mutschler2015freeze}, also induce content
exchange between giant unilamellar vesicles through
diffusion \cite{Schwille2018} or through fusion and fission \cite{Tsuji590}. 
Also, transient compartmentalization, which involves
the occasional complete mixing of contents between comparments, is
considered to be relevant to maintain functional
replicators \cite{matsuura2002importance, ichihashi2013darwinian,
  bansho2016host, matsumura2016transient, PhysRevLett.120.158101}. Taken
together, it therefore seems natural to assume that compartmentalization
is imperfect enough to allow the random exchange of fragments between
compartments at the primitive stages of life.

The model of fragmented replicases investigated above can be
conceptually compared to the hypercycle \cite{eigen_hypercycle:_1979}, a
model proposed to solve error catastrophes: Both models posit
that multiple distinct sequences are integrated into an auto-catalytic
system, which as a whole maintains a greater amount of information than
possible by a single sequence. However, the two models sharply differ in
dynamical aspects. In the fragmented replicases, the dynamics involves
the positive feedback, which biases replication toward a subset of the
fragments. In the hypercycle, the dynamics involves negative feedback,
which balances the replication of distinct sequences on a long
timescale, but also causes oscillatory instability on a short timescale. 
Given these comparisons, horizontal transfer as studied here will be also relevant to hypercycles. 
In addition, hypercycles entail evolutionary instability due
to parasites \cite{smith1979hypercycles}. 
It would be interesting to study the effect of parasites on
the fragmented ribozymes in the future.

\section*{Supporting information}

\paragraph*{Supporting text.}
\label{S1_text}
{\bf Supporting information with sections on 1) General extension of the replication to $N$-fragments ribozymes; 2) Unstable growth for small transfer rate in our simulation of compartments is due to discreteness of molecules in cells.}

\section*{Acknowledgments}
This research is partially supported by a Grant-in-Aid for Scientific Research(S) (15H05746) from the Japan Society for the Promotion of Science(JSPS).

\nolinenumbers

%
%
%


\end{document}


\beginsupplement
\maketitle

\section{General extension of the replication to $N$-fragments ribozymes}

One could straightforwardly extend the two-fragments to general $N$-fragments system, in which $N$ fragments of $X^i$ $(i=1 ,..., N)$ assemble to form the catalyst. The assembly of the catalyst from the $N$-fragments is written as
\begin{align}
X_1 + X_2 + ... + X_N \rightarrow C,   
\label{R1}
\end{align}
and the catalyst disassemble into the $N$-fragments as 
\begin{align}
C \rightarrow X_1 + X_2 + ... + X_N.    
\label{R2}
\end{align}
Each of the fragments replicates with the aid of the catalyst $C$ as
\begin{align}
X_i + C \rightarrow 2X_i + C,    
\label{R3}
\end{align}
for $i = 1,...,N$.

The dynamics of the fragments $X_i$ and the catalyst $C$ are written as
\begin{align}
\frac{dx^i}{dt} = \left( - x^1 x^2 \cdots x^N + c \right) + x^i c - x^i \phi,
\label{S1}
\end{align}
and 
\begin{align}
\frac{dc}{dt} = \left( x^1 x^2 \cdots x^N - c \right) - c \phi,
\label{S2}
\end{align}
where $x^i$ and $c$ denote the concentrations of $X_i$ and $C$, respectively. Here, all the rate constants are fixed to one. 
In the right-hand-side of the equations, the terms in the brackets denote the assembly (\ref{R1}) and the disassembly (\ref{R2}) of the catalyst, 
and the second term in Equation (\ref{S1}) denotes the replication of the fragment $X^i$ (\ref{R3}). 
Each of the last terms with $\phi$ represents dilution. 
The dilution terms with $\phi = (1-Nc)c$ are introduced to fix total number of fragments as $\sum_i (x^i + c) = \sum_i x^i + Nc = 1$.

By adding both ends of Equations (\ref{S1}) and (\ref{S2}), one can write the dynamics of total concentration of $X^i$ (total of free $X^i$ and that in the catalyst), $x^i_{tot}$ $(i = 1,...,N)$ as
\begin{align*}
\frac{dx^i_{tot}}{dt} = ( x^i_{tot} -  c )c - x^i_{tot} \phi,
\end{align*}
where the concentration of the free $X_i$, $x^i$, is written as $x^i_{tot} - c$.
Then, for arbitrary pair of $i$ and $j$ ($i, j = 1,...,N$), one obtains
\begin{align}
\frac{d}{dt} \left( \frac{x^i_{tot}}{x^j_{tot}} \right) =  \frac{c^2}{x^{j2}_{tot}} \left( x^i_{tot} - x^j_{tot} \right)
\label{generalratio}
\end{align}
Given that $(c/x^j_{tot})^2$ is positive, this indicates that a small increase of $x^i_{tot}$ over $x^j_{tot}$ grows so that the solution is unstable of balanced replication $x^i_{tot} = x^j_{tot}$ for every pair of $i$ and $j$.

\section{Unstable growth for small transfer rate in our simulation of compartments is due to discreteness of molecules in cells}

The bifurcation analysis presented in the main text explains that the system is stable for infinitely small but non-zero values of $D$ for the two subsystems. 
On the other hand, the system gets unstable in our simulation of compartments for small non-zero values of $D$, 
because the system is dominated by either of the $X$- or $Y$-dominant cells and the symmetry between $X$ and $Y$ is broken.

\begin{figure}
\includegraphics[width=\textwidth]{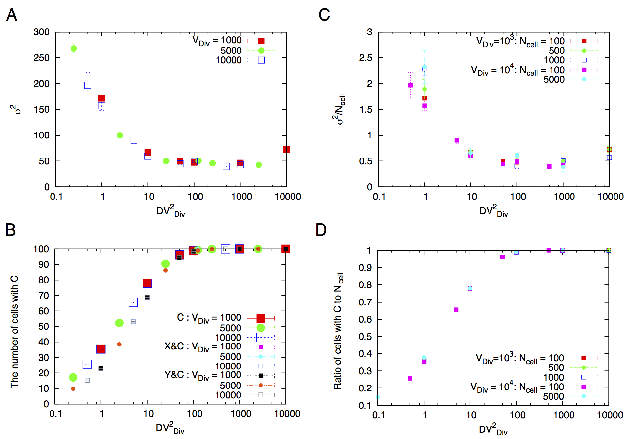}
\caption{(A) Variances $\sigma^2$ of the number of $X$-dominant cells as functions of $D$ and $V_{Div}$. Here, $N_{cell} = 100$. (B) The number of cells with catalysts $C$ scales with $DV^2_{Div}$ for $V_{Div} = 1000, 5000, 10000$. The numbers of cells with both $C$ and free $X$/$Y$ also show a similar curve. 
    Here, $N_{cell} = 100$. (C) Variances $\sigma^2$ of the number of $X$-dominant cells divided by $N_{cell}$ and (D) Ratio of cells with the catalysts $C$ to $N_{cell}$ are shown for different values of $N_{cell}$. All the points follow the same curve under the scaling $DV^2_{Div}$. }
\label{fig:S1}
\end{figure}

As $D$ decreases, the system is dominated by either of the cells because fluctuations increase between the number of $X$- and $Y$-dominant cells and, once the number of either type of cells reaches the population size $N_{cell}$, it irreversibly breaks the symmetry between $X$ and $Y$. 
Actually, the variances $\sigma^2$ of the number of $X$-dominant cells increases as $D$ decreases [Fig. \ref{fig:S1}A].

\begin{figure}
\includegraphics[width=\textwidth]{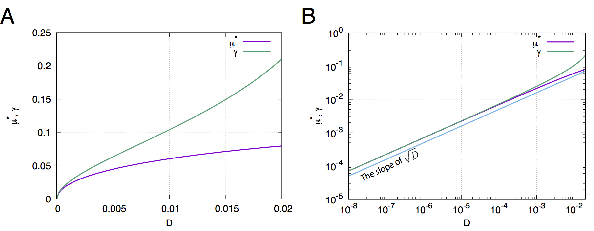}
\caption{Growth rate $\mu^*$ and $\gamma(D)$ of Equation 13 and 14 of the main text as a function of $D$ in (A) normal and (B) log-log scale for small $D$. The slope $\sqrt{D}$ is also shown in (B).}
\label{fig:S2}
\end{figure}

The increase of the variances suggests that pressure of the frequency-dependent selection to maintain the symmetry is getting weak as $D$ decreases. 
However, the analysis up to the first order of $\epsilon$ presented in the main text does not explain such dependence on $D$ for the two subsystems. 
The pressure to maintain the symmetry is represented by relative values of the factor $\gamma(D)$ to the average growth rate $\mu^*$ 
in Equations 13 and 14 of the main text.
If the relative values decrease with decreasing $D$, the pressure would be weakened because the relative increase or decrease of the growth rates, respectively, between subsystems of smaller or larger volumes gets smaller. 
Actually the factor $\gamma(D)$ depends on $D$ as $\sqrt{D}$ for small $D$ [see Fig. \ref{fig:S2}], but the average growth rate $\mu^*$ also scales as $\sqrt{D}$.
Therefore, the relative pressure of the frequency-dependent selection does not change with decreasing $D$. 

This suggests that the unstable growth in our simulation of compartments is due to the discrete nature of molecules in cells because the effect is neglected in the two subsystems.
As $D$ decreases, the number of minor fragments decreases [Fig. 3 of the main text], therefore, the cells gradually contain only a few or no catalyst.
In fact, the number of cells with the catalysts $C$ gradually decreases with decreasing $D$ [Fig. \ref{fig:S1}B: the case for both $X/Y$-fragments and catalysts
also shows similar curves] in accordance with the increase of the variances $\sigma^2$ of the number of $X$-dominant cells [Fig. \ref{fig:S1}A].
Therefore, the discrete nature of molecules in cells contributes to the enhancement of variations between cells and consequently results in the increase of the variances.

A scaling behavior between the transfer rate $D$ and the division threshold $V_{Div}$ is also consistent with the enhancement of the variances by the discreteness of molecules.
The relevant values of the transfer rate $D$ with which the effect of the discrete number of molecules appears scales as $1/V^2_{Div}$.
For substantially small $D$, the concentration of minor fragments in Equation 10 of the main text can be written as $x^1_{tot} = \sqrt{2D}$.
Therefore, the values of $D$ at which the total number of minor fragments is approximately equal to one scale as $D \approx 1/2V^2_{Div}$. 
In fact, the numbers of cells with $X$ fragments and catalysts $C$ and the variances for different $V_{Div}$ and $N_{cell}$ 
agree as one plot as a function of $DV^2_{Div}$ [Fig. \ref{fig:S1}C and D].
This suggests that, when $V_{Div}$ is infinitely large, the variances do not increase even when $D$ decreases to infinitely small. 

Further, the system is stable if the number of cells $N_{cell}$ is large because the variances of the ratio of $X$-dominant to $Y$-dominant cells 
scales as $1/\sqrt{N_{cell}}$ with $N_{cell}$. 
The variances $\sigma^2$ of the number of $X$-dominant cells divided by $N_{cell}$ collapses into the same curves for different $N_{cell}$ [Fig. \ref{fig:S1}D].
This indicates that they increases with $\sqrt{N_{cell}}$, and the ratio of $X$-dominant cells to total population scales as $1/\sqrt{N_{cell}}$.
Thus, fluctuations in the ratio of $X$-dominant to $Y$-dominant cells decreases when $N_{cell}$ is getting large.

\newpage

